\documentclass[aps,preprint,amsmath,amssymb]{revtex4}
\usepackage{graphicx}
\newcommand{\sla} {\slash\hspace{-0.25cm}}

\begin{document}
\title{\Large \bf Scalar interactions and the polarizations of
$B\to \phi K^* $}
\date{\today}
\author{\large \bf  Chuan-Hung Chen$^{a}$\footnote{Email:
phychen@mail.ncku.edu.tw} and
Chao-Qiang Geng$^{b}$\footnote{Email:
geng@phys.nthu.edu.tw} }
 \affiliation{$^{a}$Department of Physics,
National Cheng-Kung University, Tainan, 701
Taiwan\\
$^{b}$Department of Physics, National Tsing-Hua
University, Hsin-Chu , 300 Taiwan }

\begin{abstract}
We try to understand the polarization puzzle in $B\to \phi K^*$
decays with a simple Higgs model associated with flavor changing
neutral current at tree level. The new interactions can effectively
reduce the longitudinal polarization $|A_0|^2$.
In particular, we find that if
the couplings of $b$-quark in different chiralities to Higgs are the
same, the transverse polarization  $|A_\perp|^2$ can receive the largest
contribution and its value can be as large as $30\%$. On the other
hand, with opposite sign in the couplings, the other transverse
polarization $|A_{\parallel}|^2$ is enhanced.

\end{abstract}
\maketitle

In terms of naive helicity analysis, it is well known that the
transverse polarizations of vector bosons are associated with their
masses. It was expected that the partitions of vector meson
polarizations in $B$ decays should have the same behavior. As a
result, the ratio of various polarizations in two-body $B$ meson
decays can be estimated to be
\begin{eqnarray}
|A_{0}|^2 : |A_{\perp}|^2 : |A_{\parallel}|^2\sim 1:
\frac{m^2_{V}}{M^2_{B}} : \frac{m^2_{V}}{M^2_{B}}, \label{Ratio}
\end{eqnarray}
where $A_{0}$ and $A_{\parallel}$ belong to the mixtures of
S and D-wave decay amplitudes while
$A_{\perp}$ the P-wave one,
which satisfy the identity
\begin{eqnarray}
\sum_{\lambda = 0,\parallel,\perp}|A_{\lambda}|^2&=&1\,.
\label{Identity}
\end{eqnarray}
According to Eq. (\ref{Ratio}), it is believed that
in B decays with light vector mesons $A_{\perp(\parallel)}$ are much
smaller than $A_{0}$. The expectation is confirmed by BELLE
\cite{belle1} and BABAR \cite{babar1} in $B\to \rho \rho$ decays, in
which the longitudinal parts occupy over $95\%$.
 Furthermore, when the final
states include heavy vector mesons,
transverse polarizations can be relatively large.
 The conjecture is
verified in $B\to J/\Psi K^*$ decays  \cite{belle2,babar2}, in which the longitudinal contribution is
only about $ 60\%$.

However,
the rule in Eq. (\ref{Ratio}) seems  to be broken
in  $B\to \phi K^*$
decays. From the recent measurements of BELLE \cite{belle3} and BABAR
\cite{babar2,babar3}, summarized in the Table \ref{tab:pol},
it is quite clear that the longitudinal
polarizations of $B\to K^* \phi$ are only around $50\%$.
 To solve the
anomalous polarizations, the authors of Refs.
\cite{Kagan,Hou,LLNS,CCS,Li} have proposed some solutions by
introducing proper mechanisms such as large annihilation effect due
to $(S-P)\otimes (S+P)$ interactions \cite{Kagan}, the enhanced
transversality from transverse gluon emitted by $b\to s g^{(*)}$
\cite{Hou}, final state interactions \cite{LLNS,CCS} and new sets of
form factors \cite{Li}.
\begin{table}[hptb]
\caption{\label{tab:pol} The polarization fractions and relative
phases for $B\to \phi K^*$. }
\begin{ruledtabular}
\begin{tabular}{cccc}
Model &Polarization & BELLE & BABAR \\ \hline
$K^{*+}\phi$ & $|A_{0}|^2$ & $0.52 \pm 0.08 \pm 0.03$ & $0.46 \pm 0.12 \pm 0.03$\\
$$ & $|A_{\perp}|^2$ & $0.19\pm 0.08\pm 0.02$ &
$$ \\
$$ & $\phi_{\parallel}(rad)$ & $2.10\pm 0.28\pm 0.04$ &
$$ \\
$$ & $\phi_{\perp}(rad)$ & $2.31\pm 0.20 \pm 0.07$ &
$$ \\
\hline
$K^{*0} \phi$&$|A_{0}|^2$ & $0.45 \pm 0.05 \pm 0.02$ & $0.52 \pm 0.05 \pm 0.02$\\
$$ & $|A_{\perp}|^2$ & $0.30 \pm 0.06 \pm 0.02$ & $0.22\pm 0.05 \pm
0.02$\\
 $$ & $\phi_{\parallel}(rad)$ & $2.39 \pm 0.24 \pm 0.04$ &
$2.34^{+0.23}_{-0.20}\pm 0.05$ \\
$$ & $\phi_{\perp}(rad)$ & $2.51 \pm 0.23 \pm 0.04$ & $2.47\pm 0.25 \pm 0.05$
 \\
\end{tabular}
\end{ruledtabular}
\end{table}
All above proposals are related to the
uncertainities of low energy QCD. The possible new physics effects
are also studied in the literature \cite{newphys}.
In this paper, firstly we
reexamine the branching and polarization fractions of $B\to \phi K^*
$ in the framework of perturbative QCD (PQCD) by fixing the hard
scale for the involving Wilson coefficients within the SM
\cite{CKLPRD66}. And then, we introduce a new type of scalar
interactions, which allows flavor changing neutral current (FCNC) at
tree level. We will display that the
new interactions
could explain the branching ratios (BRs) and various polarizations
in $B\to \phi K^*$.

It is known that the decay amplitude of $B\to V_{1} V_{2}$
with the helicity
can be generally
parametrized as
\cite{VV}
\begin{eqnarray}
{\cal M}^{(\lambda)}
&=&\epsilon_{1\mu}^{*}(\lambda)\epsilon_{2\nu}^{*}(\lambda) \left[ a
\, g^{\mu\nu} +b\;  P_2^\mu P_1^\nu + i\, c\;
\epsilon^{\mu\nu\alpha\beta} P_{1\alpha} P_{2\beta}\right]\;.
\label{vvt}
\end{eqnarray}
Consequently, the
helicity amplitudes are
given by
\begin{eqnarray*}
H_{00}&=&\frac{-1}{2m_{1}m_{2}}\left[M^{2}_{B}-m^{2}_{1}-m^{2}_{2})a
+2M^{2}_{B} p^{2} b\right], \nonumber\\
H_{\pm\pm}&=&a\mp M_{B} p\; c, \label{helicity}
\end{eqnarray*}
where $p$ is the magnitude of vector meson momenta.
Note that we can  define the polarization amplitudes to be
\begin{eqnarray}
A_{0}=\frac{H_{00}}{(\sum_{h} |H_{h}|^2)^{1/2}}, \ \ \
A_{\parallel(\perp)}=\frac{1}{\sqrt{2} (\sum_{h} |H_{h}|^2)^{1/2}
}(H_{++} \pm H_{--}), \label{pol-amp}
\end{eqnarray}
so that Eq. (\ref{Identity}) is satisfied.
 The relative
phases between $A_{\lambda}$ are described by
$\phi_{\parallel(\perp)}=Arg(A_{\parallel(\perp)}/A_{0})$. In order
to merge the results calculated by PQCD \cite{CKLPRD66}, we rewrite
Eq. (\ref{vvt}) as
\begin{eqnarray*}
{\cal M} &=&M^{2}_{B} {\cal M}_{L} +M^{2}_{B} {\cal M}_{N}
\epsilon_{1T}^{*}\cdot\epsilon_{2T}^{*} +i {\cal M}_{T}
\varepsilon^{\alpha \beta \gamma \rho} \epsilon^{*}_{1T}
\epsilon^{*}_{2T} P_{1\gamma} P_{2\rho}.
\end{eqnarray*}
In terms of the well known effective Hamiltonian for the inclusive
$b\to s s\bar{s}$ process \cite{BBL}, the various transition matrix
elements in $B\to \phi K^* $  are written as
\begin{eqnarray}
{\cal M}_{H}=V_{tb}V^{*}_{ts} \left[ f_{\phi} E_{H}+ E^{N}_{H} +
f_{B} A_{H} + A^{N}_{H} \right],
\end{eqnarray}
where $H=L$, $N$ and $T$, $f_{\phi (B)}$ is the decay
constant of $\phi\ (B)$,  and $E_{H}^{(N)}$ and $A_{H}^{(N)}$ denote the
factorization (nonfactorization) contributions of emission and annihilation
topologies,
respectively.
 We note that the Wilson
coefficients of weak interactions have been included in $\{E\}$
and $\{A\}$ and their explicit expressions can be found in
Ref.~\cite{CKLPRD66}. For simplicity, we will fix the scale,
estimated by the energy of the exchanged hard gluon, to be
around $t=\sqrt{\bar{\Lambda}M_{B}}$ where $\bar{\Lambda}\sim
M_{B}-m_b$ with $m_b$ being the b-quark mass. The various
contributions associated with different scales are shown in
Table~\ref{tab:value_sm}.
\begin{table}[hptb]
\caption{\label{tab:value_sm} Values (in units of $10^{-3}$) of
transition matrix elements associated with different hard scales of
$t$ (GeV).}
\begin{ruledtabular}
\begin{tabular}{cccccc}
$t$  & $E_{L} $& $E^{N}_{L}$ & $A_{L} $ & $E_{N} $ & $E^{N}_{N}$ \\
\hline $2.0$&  $ -13.90$ & $0.37+i0.37$ & $1.37-8.05$ & $-2.09$ & $(-1.28+i0.04)10^{-1}$\\
$1.8$ & $-14.91$ & $0.40+i0.39$ & $1.49-i8.69$ & $-2.24$ & $(-1.41+i0.04)10^{-1}$ \\
$1.6$ & $-16.13$ & $0.43+i0.42$ & $1.62-i9.50$ & $-2.42$  & $(-1.57+i0.03)10^{-1}$\\ \hline
\hline
$t$ & $A_{N}$ & $E_{T}$ & $E^{N}_{T}$ & $A_T$ \\
\hline
$2.0$ & $-2.07+i3.15$ & $-4.08$ & $(-2.65-i0.25)10^{-1}$ & $-3.90+i6.46$ & $$\\
$1.8$ & $-2.23+i3.40$ & $-4.38$ & $(-2.92-i0.29)10^{-1}$ & $-4.22+i6.98$  \\
$1.6$ & $-2.44+i3.71$ & $-4.73$ & $(-3.26-i0.33)10^{-1}$ & $-4.62+i7.62$ \\
\end{tabular}
\end{ruledtabular}
\end{table}
Here, we have neglected the values of $A^{N}$ since  they are much
smaller than the others. In our numerical estimations, we have used
$f_{\phi}=0.237$ GeV, $f^{T}_{\phi}=0.22$ GeV, $f_{K^*}=0.22$ GeV,
$f^{T}_{K^*}=0.17$ GeV, $f_{B}=0.19$ GeV, $m_{\phi}=1.02$ GeV,
$m_{K^*}=0.89$ GeV and $M_{B}=5.28$ GeV. The wave functions of
$\phi$ and $K^*$ are refered to the results of light-cone sum rules
(LCSRs) \cite{BBKT}. Using the values of Table~\ref{tab:value_sm},
the BR and polarizations in $B_d\to \phi K^{*0}$ can be easily
obtained. To illustrate the effects of nonfactorization and
annihilation, we fix $t=1.6$ GeV and we find that $(BR, |A_{0}|^2,
|A_{\perp}|^2)=(8.1\times 10^{-6}, 0.93, 0.03)|_{E_H}$,
$(10.93\times 10^{-6}, 0.73, 0.12)|_{E_H,A_H}$ and $(9.42\times
10^{-6}, 0.62, 0.17)|_{E_H,A_H,E^{N}_{H}}$ for contributions with
$\{E_H\}$, $\{E_H,A_H\}$ and $\{E_H,A_H,E^{N}_{H}\}$, respectively.
 From the results, we find that the effects of annhilation and
nonfactorization can enhance $A_{\perp}$, but they are still not
enough to explain the central values of data in Table~\ref{tab:pol}.
 For completeness, we present the results with differnt hard
scales in Table~\ref{tab:br-po_sm}.
\begin{table}[hptb]
\caption{\label{tab:br-po_sm} Branching ratios (in units of
$10^{-6}$), polarizations and relative phases with different hard
scales of $t$ (GeV) for $B_d\to \phi K^{*0}$ in the SM.}
\begin{ruledtabular}
\begin{tabular}{ccccccc}
$t$  & $BR$& $|A_0|^2$ & $|A_{\parallel}|^2$ & $|A_{\perp}|^2 $ & $\phi_{\parallel}(rad)$ & $\phi_{\perp}(rad)$ \\
\hline $2.0$&  $ 6.93$ & $0.628$ & $0.206$ & $0.166$ & $2.16$ & $2.15$\\
$1.8$ & $8.02$ & $0.625$ & $0.207$ & $0.167$ & $2.15$ & $2.14$\\
$1.6$ & $ 9.46$ & $0.622$ & $0.209$ & $0.169$ & $2.15$ & $2.13$ \\  
\end{tabular}
\end{ruledtabular}
\end{table}
 From Table~\ref{tab:br-po_sm}, we
 note that the polarizations are
stable in different scales. It is difficult to further reduce the logitudinal polarization without
introducing new mechinism.
 As a comparsion,
 we also calculate the decay of
 $B^{+}\to \rho^{+} K^{*0}$ in the SM and, explicitly, we find that
  $(BR, |A_{0}|^2,
|A_{\perp}|^2)|_{B^{+}\to \rho^{+} K^{*0}}=(14.69 \times
10^{-6}, 0.72, 0.13)|_{E_H,A_H,E^{N}_{H}}$.
Note that the current experimental data of
 BABAR and BELLE
for $(BR,|A_{0}|^2)$ are $((17.0^{+3.5}_{-3.9})\times 10^{-6},
0.79\pm 0.09)$ \cite{babar4} and $((8.9\pm 1.7\pm
1.2)\times 10^{-6}, 0.43\pm0.11^{+0.05}_{-0.02})$ \cite{belle4},
respectively, which are not consistent with each other. Due to these
inconclusive results in $B\to\rho K^*$, in this study we
regard the polarization anomaly happens only in the decays of
$B\to \phi K^*$.

We now try to find out if there exists some
kind of new interactions which can induce large transverse
polarizations in $B\to \phi K^*$, but not in the others, such as
$B\to \rho \rho$ and $B\to \rho K^*$. Naturally, one could try the
scalar interactions in which the couplings between the scalar and
fermions are proportional to the fermion masses ($m_f$). In these
models, the down-quark pair production is expected to be one order
of magnitude smaller than that of the strange-quark pair. However,
as known, the couplings in one-Higgs-doublet and type I
two-Higgs-doublet models are suppressed by $m_{f}/m_{W}$. Although
there is an enhancement factor $\tan\beta$ in the type II Higgs
model, the effects of the $b\to s$ flavor change (FC) transition
 are one-loop suppressed. In order to get
large transverse polarizations in $B\to \phi K^*$,
 we consider a new type of scalar interactions in which FCNC at
tree level is allowed. Our another reason to try scalar interactions
is that the new contributions on transverse
polarizations should avoid the light meson mass dependence or
the power suppression of $m_{\phi}/M_{B}$, , unlike the SM in  which   $\langle \phi | \bar{s}\gamma_{\mu}s|0\rangle= m_{\phi}\epsilon_{\phi\mu}$ arises.
 For an illustration, we consider the hadronic matrix
element $\langle \phi K^* | \bar{b} s \; \bar{s} s |B\rangle$. To
get the factorizable parts, we need do the Fierz transformation.
Explicitly, we have
\begin{equation}
\langle \phi K^* | \bar{b} s \; \bar{s} s |B\rangle \propto
\frac{1}{4N_c} \langle \phi |\bar{s} \sigma_{\mu\nu} s |0\rangle\,
\langle K^*| \bar{b} \sigma^{\mu \nu} s|B\rangle + \ldots ,
\end{equation}
where $N_c=3$ is the color factor, the factor $1/4$ is from the
Fierz transformation,
$\sigma_{\mu\nu}=i[\gamma^{\mu},\gamma^{\nu}]/2$ and $\{\ldots\}$
denotes
contributions from other operators such as the vectors and axial-vectors,
which are suppressed by a factor of $m_\phi/m_B$ at the amplitude level.
 Since the nonlocal structure
of $\phi$ is related to the term $\sla\epsilon_{\phi T}
 \sla P \Phi^{T}_{\phi}$ with $\Phi^{T}_{\phi}$
being the twist-2 $\phi$ meson wave function, the factor $\langle \phi |\bar{s}
\sigma_{\mu\nu} s |0\rangle \propto \epsilon^{\mu}_{\phi} P^{\nu}
- \epsilon^{\nu}_{\phi} P^{\mu}$ which is clearly
independent of $m_\phi$.
Hence, the hadronic suppression of scalar
interactions can be only from color factor and Fierz coefficients.
Next, we will demonstrate that scalar interactions have important
influence on $B\to \phi K^*$. Before introducing a specific model,
we start from a general
interaction with a scalar boson $S$, given by
\begin{eqnarray}
{\cal L}_{eff}=\left(C_{bs} \bar{b}P_R s+ C_{sb} \bar{s} P_R b +
C_{ss} \bar{s} P_{R} s\right) S + H.c.
\label{Interaction}
\end{eqnarray}
where $P_{L(R)}=1\mp\gamma_5$. Since we are not dealing with the CP
problem, the parameters $C_{ij}$ are regarded as real numbers. From
Eq. (\ref{Interaction}) the effective interaction for the process of
$b\to s s\bar{s}$ is derived to be
\begin{eqnarray}
{\cal L}_{eff}=\frac{
C_{ss}}{m^{2}_{S}}\bar{b}\left(C_{bs}P_{R}+C_{sb}P_L\right)s\,
\bar{s} s\,,
\end{eqnarray}
where $m_{S}$ is the mass of the scalar.

\begin{figure}[htbp]
\includegraphics*[width=3.0in]{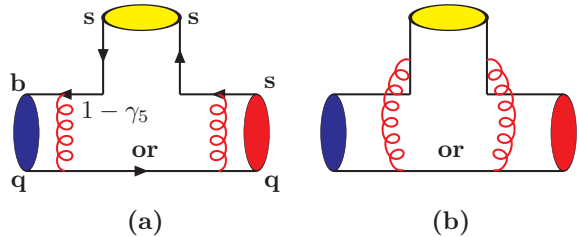} \hspace{0.3cm} \caption{
Diagrams for hadronic transition matrix elements due to the scalar
interaction of
 $\bar{b} (1-\gamma_5) s\, \bar{s} s$ with
 (a) factorizable and (b) nonfactorizable contributions.}
 \label{fig_new}
\end{figure}

The new contribution to
$B\to \phi K^*$ due to the scalar interaction are shown in Fig.~\ref{fig_new}, where
(a) and (b) stand for the factorizable and
nonfactorizable effects, respectively. Since we have assumed that the couplings
of the scalar interaction to fermions are proportional to $m_f$,
the
annihilation topologies can be neglected due to the suppression
of $m_u/v$ or $m_d/v$, comparing to emission topologies (Fig.~\ref{fig_new}) associated with $m_s/v$.
Similar to the SM case, the decay amplitudes for various
helicities could be written as
\begin{eqnarray}
{\cal M}^{NP}_{H}=\frac{C_{ss} C_{sb} }{2N_{c}m^{2}_{S}}\left[
f_{\phi} F_{H}+ N_{H} \right],
\end{eqnarray}
where $F_{H}$ and $N_{H}$ are the factorizable and nonfactorizable
effects, respectively.  Here, for simplicity, we have only presented
the contributions of $C_{bs}$. The result of $C_{sb}$ can be
obtained by changing the sign in $H=L$ and $N$. $C_{sb}$ and
$C_{bs}$ have the same contributions for $H=T$. The explicit
expressions of $F_{H}$ and $N_{H}$ are shown in
Appendix~\ref{hardamp} and their values are given in
Table~\ref{tab:value_sm}.
\begin{table}[hptb]
\caption{\label{tab:value_new} Values (in units of $10^{-2}$) of
transition matrix elements for scalar interactions.}
\begin{ruledtabular}
\begin{tabular}{ccccccc}
 & $F_{L}$ & $N_{L}$ & $F_{N} $ & $N_{N} $ & $F_T$ & $F_N$ \\
\hline &  $ 9.74$ & $-2.85+i0.26$ & $-6.91$ & $0.44+i0.31$ & $-20.27$ & $-0.038+i0.29$\\
\end{tabular}
\end{ruledtabular}
\end{table}

For a specific model, we concentrate on the generalized
two-Higgs-doublet model (Model III) and the corresponding Yukawa
Lagrangian for down-type quarks is described by \cite{ARSPRD55}
\begin{eqnarray}
{\cal L}^{(III)}_{Y}=\eta^{D}_{ij} \bar{Q}_{iL} \Phi_1 D_{jR} +
\xi^{D}_{ij} \bar{Q}_{iL} \Phi_2 D_{jR}  +H.c.\; ,
\end{eqnarray}
where the indices $i (j)$ represent the possible quark flavors and
$\xi^{D}_{ij}$ denote the allowed FC effects. The vacuum expectation
values (VEVs) of neutral Higgs fields are denoted by $\langle
\Phi^{0}_{1(2)} \rangle=v_{1(2)}$. For convenience,  we can choose a
proper basis such that only one scalar field possesses the VEV.
Hence, the new scalar fields could be chosen to be
$\phi^{0}_{1}=\cos\beta\Phi^{0}_{1}+\sin\beta \Phi^{0}_{2}=(v+ H^{0}
+i \chi^{0})/\sqrt{2}$ and
$\phi^{0}_{2}=-\sin\beta\Phi^{0}_{1}+\cos\beta \Phi^{0}_{2}=(H^{1}+
i H^{2})/\sqrt{2}$, where $v=\sqrt{v^2_{1}+v^{2}_{2}}$ is the
vacuum expectation value (VEV) of $\phi^{0}_1$, $\cos\beta
(\sin\beta)=v_{1(2)}/v$, $H^{0(1)}$ and $H^{2}$  are CP-even and
CP-odd Higgs bosons,
 and $\chi^0$ is Goldstone boson, respectively.
Since $H^0$ and $H^1$ are not physical eigenstates, the mass
eigenstates could be parametrized by a mixing angle $\alpha$ as
$h^{0}_{SM}=H^0 \cos\alpha + H^1 \sin\alpha$ and $h^{0}=-H^0
\sin\alpha + H^1 \cos\alpha$. When $\alpha$ goes to zero, $h^0_{SM}$
becomes the SM Higgs. It is known that to get naturally small FCNC
at tree level, one can use the ansatz \cite{ARSPRD55,CS}
\begin{eqnarray}
\xi^{D}_{ij}=\lambda_{ij} \frac{\sqrt{m_im_j }}{v}.
\end{eqnarray}
It has been analyzed phenomenologically that the coupling
$\lambda_{sb}$ for the transition of $b\to s$ may not be small and
it could be as large as  $O(10)$ \cite{ARSPRD55}. Besides the
coupling $\lambda_{sb}$, for $b\to s s \bar{s}$, we also need the
information on $\lambda_{ss}$, which is flavor conserved. To
understand the order of magnitude on $\lambda_{ss}$, we refer to the
case of type II model, in which the corresponding coupling $\bar{s}
s H^{1}$ is $m_{s}\cos\alpha/(v\cos\beta)$, $i.e.$, $\lambda_{ss}$
is order of $\cos\alpha/\cos\beta$. In the scenario of a large
$\tan\beta=v_{2}/v_{1}$, $\lambda_{ss}$ could be order of
$\tan\beta\sim m_{t}/m_b$.
We note that the large enhancement of $\lambda_{ss}$ is natural only
for the type II model. In our considered type III model, we shall set
$\lambda_{ss}=O(100)$, which implies that the Higgs coupling to the strange quark $\xi^D_{ss}=\lambda_{ss}m_{s}/v$ is $O(10^{-2})$.
Since we only concern the non-leptonic decays, it is clear that
the value of $\lambda_{\mu\mu}=O(1)$ for the muonic coupling given
in Ref. \cite{ARSPRD55} can be relaxed and there are no stringent limits for
$\lambda_{ij}$.

\begin{figure}[htbp]
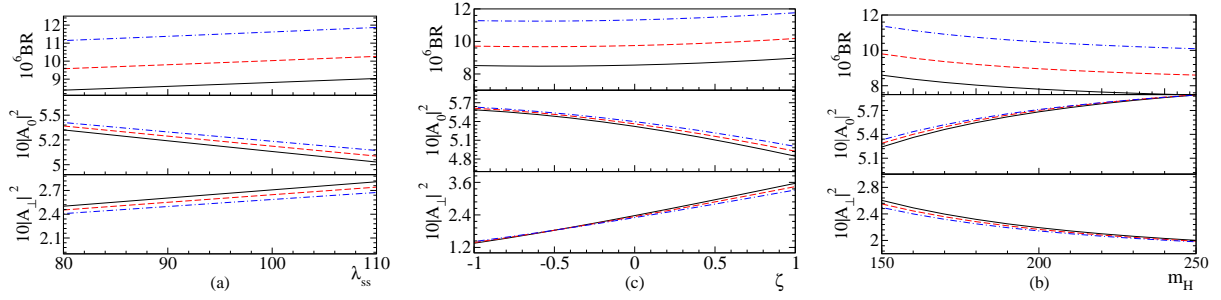

\includegraphics[width=2.0in]{brpo_lss} \hspace{0.1cm} \includegraphics[width=2.0in]{brpo_zeta}
\hspace{0.1cm} \includegraphics[width=2.0in]{brpo_mh} \caption{
Branching ratio and $|A_{\parallel (\perp)}|^2$ in $B_d\to \phi
K^{*0}$ as functions of (a) $\lambda_{ss}$ with $\zeta=0.2$ and
$m_{H}=150$ GeV, (b) $\zeta$ with $\lambda_{ss}=90$ and $m_{H}=150$
GeV and  (c) $m_{H}$ with $\zeta=0.2$ and $\lambda_{ss}=90$,
respectively. The solid, dashed and dotted curves stand for $t=2.0$,
 $1.8$ and $1.6$ GeV, respectively.}
 \label{fig:brpo_new}
\end{figure}

To estimate the influence of scalar interactions, we set
$\lambda_{ss}=90$, $\lambda_{sb}=5$ and $m_{H}=150$ GeV and take
$\zeta=\lambda_{bs}/\lambda_{sb}$ as a variable. By using the results
of Tables~\ref{tab:value_sm}
and \ref{tab:value_new}, we present the BR and polarizations of
$B_d\to \phi K^{*0}$ for different values of $\zeta$ in
Table~\ref{tab:br-po_new}. In Fig.~\ref{fig:brpo_new}, we show BR
and $|A_{\parallel (\perp)}|^2 $ as functions of (a) $\lambda_{ss}$
with $\zeta=0.2$ and $m_{H}=150$ GeV, (b) $\zeta$ with
$\lambda_{ss}=90$ and $m_{H}=150$ GeV and  (c) $m_{H}$ with
$\zeta=0.2$ and $\lambda_{ss}=90$, respectively.
\begin{table}[hptb]
\caption{\label{tab:br-po_new} Branching ratios (in units of
$10^{-6}$), polarizations and relative phases  of $B_d\to \phi
K^{*0}$ by combining the results of Tables~\ref{tab:value_sm}
and
\ref{tab:value_new} with $t=1.6$ GeV, $m_{H}=150$ GeV, $\lambda_{ss}=90$ and  $\lambda_{sb}=5$.
}
\begin{ruledtabular}
\begin{tabular}{ccccccc}
$\zeta$  & $BR$& $|A_0|^2$ & $|A_{\parallel}|^2$ & $|A_{\perp}|^2 $ & $\phi_{\parallel}(rad)$ &  $\phi_{\perp}(rad)$ \\
\hline $-1.0$ & $11.28$ & $0.56$ & $0.30$ & $0.14$& $2.16$ & $2.18$\\
$-0.6$ & $11.25$ & $0.56$ & $0.27$ & $0.17$ & $2.24$ & $2.23$\\
$-0.2$ & $11.28$ & $0.55$ & $0.24$ & $0.21$ & $2.22$ & $2.26$ \\
$0.0$&  $11.32$ & $0.54$ & $0.23$ & $0.23$ & $2.21$ & $2.27$\\
$0.2$ & $11.37$ & $0.53$ & $0.22$ & $0.25$ & $2.20$ & $2.28$\\
$0.6$ & $11.53$ & $0.52$ & $0.19$ & $0.29$ & $2.17$ & $2.30$ \\
$1.0$ & $11.76$ & $0.50$ & $0.17$ & $0.33$ & $2.15$ & $2.31$ \\
\end{tabular}
\end{ruledtabular}
\end{table}
 From the results of Table~\ref{tab:br-po_new} and
 Fig.~\ref{fig:brpo_new}, we find that
 $|A_{\perp}|^2$ increases (decreases) if
$\zeta>0\ (<0)$. In particular,
when $\zeta> 0.6$,
 $|A_{\perp}|^2$ can be as large as $30\%$.
 We remark that
 the contributions of $\lambda_{sb}$ and $\lambda_{bs}$ to ${\cal
 M}_{L,N}$
 are opposite in sign but to ${\cal M}_{T}$ the
 same sign. Therefore, if we take $\lambda_{sb}=\lambda_{bs}$, the
 scalar interactions can only contribute to $|A_{\perp}|^2$.
 On the other hand, if
 $\lambda_{sb}=-\lambda_{bs}$, only $|A_{\parallel}|^2$ gets affected
as shown in Table~\ref{tab:br-po_new}.

In summary, we have studied how scalar interactions effectively
affect the polarizations in $B\to \phi K^*$ decays. We have illustrated that for
the scalar interactions with FCNC couplings of $C_{sb}=C_{bs}\sim 1.6\times 10^{-2}$ and  FC one of $C_{ss}\sim 4.7\times 10^{-2}$,
the longitudinal polarization $|A_{0}|^2$ can be $50 \%$ and
the transverse polarization $|A_{\perp}|^2$  $30 \%$. We have also found
that the sign of $\zeta=\lambda_{sb}/\lambda_{bs}=C_{sb}/C_{bs}$
controls the relative magnitudes of $A_{\parallel}$ and
$A_{\perp}$.\\

{\bf Acknowledgments}\\

This work is supported in part by the National Science Council of
R.O.C. under Grant \#s: NSC-93-2112-M-006-010 and
NSC-93-2112-M-007-014.

\section{Appendix: Decay amplitudes for new scalar interactions}
\label{hardamp}

The transition matrix elements of factorizable and nonfactorizable
effects for the effective interaction $\bar{b}(1-\gamma_5)s\, \bar{s} s$ are
given as follows: the factorizable amplitudes with various
helicities are
\begin{eqnarray}
F_{L}&=&2\pi C_F M^{2}_{B}\int^{1}_{0}dx_{1} dx_{3}
\int^{\infty}_{0} b_{1} db_{1} b_{3} db_{3} \Phi_{B}(x_{1}) \{
[(1+x_{3}-r_{\phi}) \Phi_{K^*}(x_3)\nonumber
\\
&&
+r_{K^*}(1-2x_3+r_{\phi}x_{3})\left(\Phi^{t}(x_{3})+\Phi^{s}_{K^*}(x_3)\right)]
E_{e}(t^{(1)}_e) h_{e}(x_1,x_3,b_1,b_3)\nonumber \\
&&
+[2r_{K^*}(1-r_{\phi})\Phi^{s}_{K^*}(x_3)]E_e(t^{(2)}_e)h_{e}(x_3,x_1,b_3,b_1)\},
\end{eqnarray}
\begin{eqnarray}
F_{N}&=&-2\pi C_F M^{2}_{B}\int^{1}_{0}dx_{1} dx_{3}
\int^{\infty}_{0} b_{1} db_{1} b_{3} db_{3} \Phi_{B}(x_{1}) \{
[(1+x_{3}-r_{\phi}) \Phi^{T}_{K^*}(x_3)\nonumber
\\
&&
+r_{K^*}(1-2x_3)\left(\Phi^{v}(x_{3})+\Phi^{a}_{K^*}(x_3)\right)
-r_{\phi}r_{K^*}\left((2+x_3)\Phi^{v}_{K^*}(x_3)-x_3\Phi^{a}_{K^*}(x_3)
\right)]
\nonumber \\
&& \times E_e(t^{(1)}_e) h_{e}(x_1,x_3,b_1,b_3)
+[r_{K^*}(1-r_{\phi})\left(\Phi^{v}_{K^*}(x_3)+\Phi^{a}_{K^*}(x_3)\right)]\nonumber \\
&& \times E_e(t^{(2)}_e)h_{e}(x_3,x_1,b_3,b_1)\},
\end{eqnarray}
\begin{eqnarray}
F_{T}&=&-4\pi C_F M^{2}_{B}\int^{1}_{0}dx_{1} dx_{3}
\int^{\infty}_{0} b_{1} db_{1} b_{3} db_{3} \Phi_{B}(x_{1}) \{
[(1+x_{3}+r_{\phi}) \Phi^{T}_{K^*}(x_3)\nonumber
\\
&&
+r_{K^*}(1-2x_3)\left(\Phi^{v}(x_{3})+\Phi^{a}_{K^*}(x_3)\right)
-r_{\phi}r_{K^*}\left((2+x_3)\Phi^{a}_{K^*}(x_3)-x_3\Phi^{v}_{K^*}(x_3)
\right)]
\nonumber \\
&& \times E_e(t^{(1)}_e) h_{e}(x_1,x_3,b_1,b_3)
+[r_{K^*}(1-r_{\phi})\left(\Phi^{v}_{K^*}(x_3)+\Phi^{a}_{K^*}(x_3)\right)]
\nonumber \\ && \times E_e(t^{(2)}_e)h_{e}(x_3,x_1,b_3,b_1)\},
\end{eqnarray}
where $\{\Phi\}$ denote the distribution amplitudes of $\phi$ and
$K^*$ mesons. We consider the effects up to twist-3. The hard
function $h_{e}$ and $E_e$ factor  are
\begin{eqnarray*}
h_{e}(x_1,x_3,b_1,b_3)&=&K_{0}(\sqrt{x_1
x_3}M_{b}b_1)S_{t}(x_3)[\theta(b_1-b_3)K_0(\sqrt{x_3}M_B b_1)
I_{0}(\sqrt{x_3}M_B b_3)\nonumber \\&&+\theta(b_3-b_1)
K_0(\sqrt{x_3}M_B b_3)I_0(\sqrt{x_3}M_B b_1)], \\
E_{e}(t)&=&\alpha_{s}(t)S_{B}(t)S_{K^*}(t).
\end{eqnarray*}
The Sudakov factors for $K^*$ and $B$ mesons and threshold
resummation factor are given by
\begin{eqnarray*}
S_{B}&=&exp\left[ -s(x_1 P^{+}_{B},b_1)-2\int^{t}_{1/b_1} \frac{d
\mu}{\mu} \gamma(\alpha_{s}(\mu))\right], \\
S_{K^*}&=&exp\left[ -s(x_3
P^{+}_{3},b_3)-s((1-x_3)P^{+}_{2},b_3)-2\int^{t}_{1/b_2} \frac{d
\mu}{\mu} \gamma(\alpha_{s}(\mu))\right], \\
 S_{t}(x)&=&{2^{1+2c}\Gamma(3/2+c)\over \sqrt{\pi}\Gamma(1+c)
 [x(1-x)]^{c}},
\end{eqnarray*}
where $\gamma=-\alpha_s/\pi$ which is the quark anomalous dimension,
the variables $(b_{1}, b_2,b_3)$ are conjugate to the parton
transverse momenta $(k_{1T},k_{2T},k_{3T})$, $c=0.4$ for $B\to \phi
K^*$ decays, and the explicit expression for $s(x,b)$ can be found
in Ref. \cite{BS}. The scale $t^{(1)}_{e}$ and $t^{(2)}_{e}$ are
chosen by
\begin{eqnarray*}
t^{(1)}_{e} &=&max(\sqrt{x_3}M_{B},1/b_1,1/b_3), \\
t^{(2)}_{e}&=&max(\sqrt{x_1}M_{B},1/b_1,1/b_3).
\end{eqnarray*}
The nonfactorizable amplitudes with various helicities are given as
\begin{eqnarray}
N_{L}&=&-4\pi C_F M^{2}_{B}\sqrt{2N_c}\int^{1}_{0}d[x]
\int^{\infty}_{0} b_{1} db_{1} b_{2} db_{2} \Phi_{B}(x_{1}) \{
[x_{2} \Phi_{\phi}(x_2)\Phi_{K^*}(x_3)\nonumber \\
&& +r_{K^*} x_3 \left( \Phi^{t}(x_{3})-\Phi^{s}_{K^*}(x_3)\right)
-r_{\phi}r_{K^*}\Phi^{t}_{\phi}(x_2)\left(
0.5(x_2+3x_3)\Phi^{t}_{K^*}(x_3)+x_2 \Phi^{s}_{K^*}(x_3)\right)
\nonumber \\&&+2r_{\phi}x_2 \Phi^{t}_{\phi}(x_2) \Phi_{K^*}(x_3)
-r_{\phi}r_{K^*}\left( (x_2-x_3)\Phi^{t}_{K^*}(x_3)+(x_2+x_3)
\Phi^{s}_{K^*}(x_3)\right)]\nonumber
\\ && \times E_d(t^{(1)}_d)h^{(1)}_{d}(x_1,x_2,x_3,b_1,b_2)
-[2(1-x_2+x_{3}) \Phi_{\phi}(x_2)\Phi_{K^*}(x_3)\nonumber \\&&
+r_{K^*}x_{3}\Phi_{\phi}(x_2)\left(\Phi^{t}(x_{3})+\Phi^{s}_{K^*}(x_3)\right)
 -r_{\phi}(1-x_2)\left(\Phi^{t}_{\phi}(x_2)-\Phi^{s}_{\phi}(x_2)
\right)\Phi_{K^*}(x_3)\nonumber
\\&&
+2r_{\phi}r_{K^*}(1-x_2+x_3)\left(\Phi^{t}_{\phi}(x_2)\Phi^{t}_{K^*}(x_3)
-\Phi^{s}_{\phi}(x_2)\Phi^{s}_{K^*}(x_3) \right)] \nonumber
\\ && \times E_d(t^{(2)}_{d}) h^{(1)}_{d}(x_1,x_2,x_3,b_1,b_2)\},
\end{eqnarray}
\begin{eqnarray}
N_{N}&=&-4\pi C_F M^{2}_{B}\sqrt{2N_c}\int^{1}_{0}d[x]
\int^{\infty}_{0} b_{1} db_{1} b_{2} db_{2} \Phi_{B}(x_{1}) \{
[-x_2 \Phi^{T}_{\phi}(x_2)\Phi^{T}_{K^*}(x_3)\nonumber
\\
&&+r_{\phi}x_2\left(
\Phi^{v}_{\phi}(x_2)-\Phi^{a}_{\phi}(x_2)\right)\Phi^{T}_{K^*}(x_3)]
E_d(t^{(1)}_{d})h^{(1)}_d(x_1,x_2,x_3,b_1,b_2) \nonumber \\
&& +[(1-x_2+x_3)\Phi^{T}_{\phi}(x_2)
\Phi^{T}_{K^*}(x_3)-r_{K^*}x_3 \Phi^{T}_{\phi}(x_2) \left(
\Phi^{v}_{K^*}(x_3)+\Phi^{a}_{K^*}(x_3)\right)\nonumber \\
&& +r_{\phi}(1-x_2)\left(
\Phi^{v}_{\phi}(x_2)-\Phi^{a}_{\phi}(x_2)\right)\Phi^{T}_{K^*}(x_3)-2r_{\phi}r_{K^*}
(1-x_2+x_3)\nonumber \\ && \times\left(\Phi^{v}_{\phi}(x_2)
\Phi^{v}_{K^*}(x_3)-\Phi^{a}_{\phi}(x_2) \Phi^{a}_{K^*}(x_3)
\right)]E_d(t^{(2)}_{d}) h^{(2)}_{d}(x_1,x_2,x_3,b_1,b_2),
\end{eqnarray}
\begin{eqnarray}
N_{T}&=&8\pi C_F M^{2}_{B}\sqrt{2N_c}\int^{1}_{0}d[x]
\int^{\infty}_{0} b_{1} db_{1} b_{2} db_{2} \Phi_{B}(x_{1}) \{
[x_2 \Phi^{T}_{\phi}(x_2)\Phi^{T}_{K^*}(x_3)\nonumber
\\
&&+r_{\phi}x_2\left(
\Phi^{v}_{\phi}(x_2)-\Phi^{a}_{\phi}(x_2)\right)\Phi^{T}_{K^*}(x_3)]
E_d(t^{(1)}_{d})h^{(1)}_d(x_1,x_2,x_3,b_1,b_2) \nonumber \\
&& -[(1-x_2+x_3)\Phi^{T}_{\phi}(x_2)
\Phi^{T}_{K^*}(x_3)-r_{K^*}x_3 \Phi^{T}_{\phi}(x_2) \left(
\Phi^{v}_{K^*}(x_3)+\Phi^{a}_{K^*}(x_3)\right)\nonumber \\
&& -r_{\phi}(1-x_2)\left(
\Phi^{v}_{\phi}(x_2)-\Phi^{a}_{\phi}(x_2)\right)\Phi^{T}_{K^*}(x_3)+2r_{\phi}r_{K^*}
(1-x_2+x_3)\nonumber \\ && \times\left(\Phi^{v}_{\phi}(x_2)
\Phi^{a}_{K^*}(x_3)-\Phi^{a}_{\phi}(x_2) \Phi^{v}_{K^*}(x_3)
\right)]E_d(t^{(2)}_{d}) h^{(2)}_{d}(x_1,x_2,x_3,b_1,b_2),
\end{eqnarray}
where the Sudakov factor for the $\phi$ meson is given as
\begin{eqnarray*}
S_{\phi}&=&exp\left[ -s(x_2
P^{+}_{2},b_2)-s((1-x_2)P^{+}_{2},b_2)-2\int^{t}_{1/b_2} \frac{d
\mu}{\mu} \gamma(\alpha_{s}(\mu))\right],
\end{eqnarray*}
and the hard functions $h^{(j)}_d$ are
\begin{eqnarray*}
h^{(j)}_{d}&=& [\theta(b_1-b_2) K_{0}(DM_{B}b_1) I_{0}(DM_{B}b_2)
+\theta(b_2-b_1)K_0(DM_B b_2)I_{0}(DM_{B} b_1)] \nonumber \\
 &&\times \left\{
  \begin{array}{c}
    K_{0}(D_{j}M_{B}b_2) \ \ \ \ \ \ \ \ {\rm for\ D^{2}_{j}\geq 0}, \\
     {i\pi \over 2} H^{(1)}_{0}(\sqrt{|D^2_j|}M_B b_2)\ \ \ {\rm for\ D^{2}_{j}\leq
     0}
                                                  \end{array}
                                                \right.
\end{eqnarray*}
with $D^2=x_1 x_3$, $D^2_1=(x_1-x_2)x_3$ and
$D^2_2=-(1-x_1-x_2)x_3$. The scales $t^{(j)}_{d}$ are chosen by
\begin{eqnarray*}
t^{(1)}_{d}&=&max(DM_B,\sqrt{|D^2_1|}M_B,1/b_1,1/b_2), \\
t^{(2)}_{d}&=&max(DM_B,\sqrt{|D^2_2|}M_B,1/b_1,1/b_2).
\end{eqnarray*}



\begin{thebibliography}{99}
\bibitem{belle1}BELLE Collaboration, J. Zhang {\it et al.} , Phys. Rev. Lett. 91,
221801 (2003).

\bibitem{babar1} BABAR Collaboration, B. Aubert {\it et al.}, Phys. Rev. Lett.
91, 171802 (2003).

\bibitem{belle2} BELLE Collaboration, K. Abe {\it et al.}, Phys. Lett. B{\bf 538}, 11 (2002); hep-ex/0408104.

\bibitem{babar2} BABAR Collaboration, B. Aubert {\it et al.},
Phys. Rev. Lett. {\bf 87}, 241801 (2001).

\bibitem{belle3}BELLE Collaboration, K. F. Chen, {\it et al.}, hep-ex/0503013.

\bibitem{babar3}BABAR Collaboration, B. Aubert {\it et al.}, Phys. Rev. Lett. {\bf 93}, 231804 (2004).

\bibitem{Kagan} A. Kagan, Phys. Lett. B{\bf 601}, 151 (2004).

\bibitem{Hou} W.S. Hou and M. Nagashima, hep-ph/0408007.

\bibitem{LLNS} P. Colangelo, F. De Fazio and T.N. Pham, Phys. Lett. B{\bf 597}, 291 (2004);
 M. Ladisa {\it et al.},  Phys. Rev. D{\bf 70}, 114025 (2004).

\bibitem{CCS}H.Y. Cheng, C.K. Chua and A. Soni, Phys. Rev. D{\bf 71}, 014030
(2005).

\bibitem{Li} H.N. Li, hep-ph/0411305.

\bibitem{newphys} A. Kagan, hep-ph/0407076; E. Alvarez {\it et al}, Phys. Rev. D{\bf 70},
 115014 (2004);
Y.D. Yang, R.M. Wang and G.R. Lu, hep-ph/0411211; A.K. Giri and R.
Mohanta, hep-ph/0412107; P.K. Das and K.C. Yang, Phys. Rev. D{\bf
71}, 094002 (2005); C.S. Kim and Y.D. Yang, hep-ph/0412364.

\bibitem{CKLPRD66} C.H. Chen, Y.Y. Keum and H.N. Li, Phys. Rev.
D{\bf 66}, 054013 (2002).

\bibitem{VV} G. Valencia, Phys. Rev. D{\bf 39}, 3339 (1989); G.
Kramer and W.F. Paimer, Phys. Rev. D{\bf 45}, 193 (1992).


\bibitem{BBL} G. Buchalla, A.J. Buras and M.E. Lautenbacher, Rev.
Mod. Phys. {\bf 68}, 1125 (1996).

\bibitem{BBKT} P. Ball {\it et al.}, Nucl. Phys. B{\bf 529}, 323
(1998).

\bibitem{babar4} BABAR Collaboration, B. Aubert {\it et al},
hep-ex/0408093.

\bibitem{belle4} BELLE Collaboration, J. Zhang {\it et al.},
hep-ex/0505039.

\bibitem{ARSPRD55} D. Atwood, L. Reina and A. Soni, Phys. Rev.
D{\bf 55}, 3156 (1997).

\bibitem{CS}T.P. Cheng and M. Sher, Phys. Rev. D{\bf 35}, 3484
(1987); {\it ibid} {\bf 44}, 1461 (1991).

\bibitem{BS}J. Botts and G. Sterman, Nucl. Phys. B{\bf 325}, 62
(1989); H.N. Li and G. Sterman, Nucl. Phys. B{\bf 381}, 129 (1992).


\end{thebibliography}
\end{document}